\documentclass[review,12pt]{elsarticle}

\usepackage{geometry}

\usepackage{color,soul}
\usepackage{adjustbox}
\usepackage{graphicx} 
\usepackage{caption}
\usepackage{subcaption}

\usepackage{float}
\usepackage{amssymb}
\usepackage{amsmath}
\usepackage{amsthm}
\usepackage{lineno}

\usepackage{nomencl}
\makenomenclature
\usepackage{etoolbox}
\renewcommand\nomgroup[1]{%
  \item[\bfseries
  \ifstrequal{#1}{G}{Greek Symbols}{%
  \ifstrequal{#1}{S}{Superscripts and subscripts}{}}%
]}

\journal{International Journal of Heat and Mass Transfer}

\begin{document}
\nolinenumbers
\begin{frontmatter}



\title{Analytical and numerical solutions to the non-diffusive Stefan problem}


\author[label1]{Matthew Van Ham}
\author[label3]{Minghan Xu} 
\author[label1,label2]{Samuel Huberman}
\ead{samuel.huberman@mcgill.ca}
\affiliation[label1]{organization={Department of Chemical Engineering, McGill University},
            addressline={3610 University Street}, 
            city={Montreal},
            postcode={H3A 0C5}, 
            state={QC},
            country={Canada}}

\affiliation[label3]{organization={Department of Civil and Mineral Engineering, University of Toronto},
            addressline={35 St.~George Street}, 
            city={Toronto},
            postcode={M5S 1A4}, 
            state={ON},
            country={Canada}}

\affiliation[label2]{organization={Department of Physics, McGill University},
            addressline={3600 University Street}, 
            city={Montreal},
            postcode={H3A 2T8}, 
            state={QC},
            country={Canada}}
\begin{abstract}
In this work, the Maxwell--Cattaneo--Vernotte (MCV) equation is used to model the one-dimensional hyperbolic Stefan problem in the limit of a small Stefan number (Ste $\ll$ 1). The solutions are approximated with perturbation series expansions using a reformulation in which time is expressed as a function of the solid-liquid interface position. The first proposed solution is derived in a framework that considers diffusive heat transfer at the phase change interface, for analytic tractability. Two rectification strategies are proposed to address the asymptotic divergence present in this formulation: a rescaled inner solution which is then combined with the outer solution to yield a composite solution, and size-dependent thermo-physical system parameters for better capture of hyperbolic effects at the phase change interface. The resulting interface profiles exhibit a characteristic parabolic-like shape, consistent with diffusive Stefan problem findings, with pronounced early-time hyperbolic effects at larger thermal relaxation times. Parametric studies are done over three pertinent variables in the dimensionless system: the Stefan number ($\mathrm{Ste}$), the dimensionless thermal relaxation time ($\widetilde \tau$), and the thermal diffusivity ($\alpha$). The studies suggest that model error scales with the Stefan number in accordance with the theoretical truncation error of the perturbation expansion. Additionally, larger values of $\widetilde \tau$ amplify early-time hyperbolic effects, while larger $\alpha$ extends the relative temporal domain over which these hyperbolic effects remain significant; both result in increased model error.
\\
\end{abstract}

\begin{highlights}
\item Developed perturbation series solutions to the hyperbolic Stefan problem.
\item Developed a composite solution profile which resolves early-time wave-like thermal effects.
\item Incorporated size-dependent thermophysical properties to account for hyperbolic premelting effects.
\item Outlined a TVD-MacCormack solver with numerically selective artificial viscosity.
\end{highlights}


\end{frontmatter}



\section{Introduction}
Solid-liquid phase change with moving boundary dynamics is prominent in a wide range of engineering applications and natural systems, including phase change materials (PCMs) \cite{alexiades1993mathematical,crank1984free,elsayed2015phase,regin2008heat}, permafrost evolution dynamics \cite{zhu2012stefan,woo2012permafrost,kurylyk2013mathematical,rasmussen2021method}, droplet freezing \cite{shikhmurzaev2007diffusion,kornev1983freezing,davis2001theory,wettlaufer1999ice}, and microscale melting in materials \cite{tzou1997macrotomicroscale,cattaneo1958form,joseph1989heatwaves,samarskii1995computational}. The underlying mathematical model is typically treated as a Stefan problem \cite{stefan1891}
, one of the most widely studied free boundary problems, where classical heat equations, applied in each phase, are coupled with a moving interfacial flux boundary condition. Under classical Fourier heat conduction, the temperature profile within the domain is parabolic, reflecting diffusive heat transport with an infinite propagation speed and therefore a fundamentally local thermal response. This framework has been studied extensively, yielding well-developed closed form analytic solutions from similarity solutions and perturbation expansions \cite{ceretani2013similarity, xu2020development}.

Fourier's law breaks down when thermal relaxation effects operate on a similar characteristic time-scale as the transient problem being modeled \cite{joseph1989heatwaves}. Under these conditions, the instantaneous information wave propagation, assumed within a classic diffusive framework, is no longer valid. Instead, the heat transfer exhibits nonlocal behavior, in which the temperature field evolves in response to propagating thermal disturbances rather than purely diffusive entropy. As a result, the solution to the governing partial differential equations (PDEs) at any point depends on the position of the thermal and phase change wave fronts, moving through the spatial domain with a finite characteristic speed. Such behavior is observed in  nanoscale systems, low-temperature solids, ultra fast laser heating, and materials exhibiting phonon hydrodynamics or ballistic transport \cite{chen2005nanoscale,huberman2019secondsound,kovacs2018analytic,volz2007microscale}. In these conditions, hyperbolic modifications to the diffusive heat equation have been developed to capture of a thermal wave-front by incorporating a finite thermal relaxation time \cite{jitendra2021numerical,mabood2021dynamics,kovacs2019numerical,hennessy2019modelling}.

Although hyperbolic heat transfer has been extensively studied in previous literature \cite{hennessy2018asymptotic,vernotte1958paradoxes,zhu2012stefan}, relatively little attention has been devoted to studying hyperbolic effects within a free boundary problem.  Incorporating finite-speed, wave-like thermal effects into the Stefan framework introduces additional nonlinear coupling between the temperature field and the moving interface, significantly increasing both analytical and numerical complexity \cite{glass1990hyperbolic,glass1991formulation,belhamadia2023numerical}. The increase in complexity arises from the change in character of the governing PDE: the introduction of a finite thermal relaxation time gives rise to a second-order time derivative, rendering the equation hyperbolic rather than parabolic. The hyperbolic formulation becomes increasingly difficult to solve because it preserves singularities and propagates information at finite speeds along specific characteristic curves, rather than instantly smoothing data as in the parabolic case. This in turn introduces sharper gradients and wave-fronts that require more intensive numerical schemes or analytic formulations to approximate the behavior of the phase change. On the numerical side, a substantial body of work has developed high-order solvers for the hyperbolic heat equation \cite{glass1990hyperbolic,glass1991formulation,belhamadia2023numerical,tzou1997macrotomicroscale}, though analytical modelling of hyperbolic moving boundary problems has received comparatively less attention. The differences of hyperbolic and parabolic temperature profiles in a moving boundary problem are illustrated in Figure 1, a snapshot in time of both hyperbolic and parabolic numerical simulations with the domain approximately 40 percent frozen. The principal numerical difficulty in solving the hyperbolic Stefan problem arises from the discontinuous jump in temperature across the solid–liquid interface, which can lead to Gibbs-type oscillations if not properly treated.

\begin{figure}[H]
    \centering
    \includegraphics[width=0.75\textwidth]{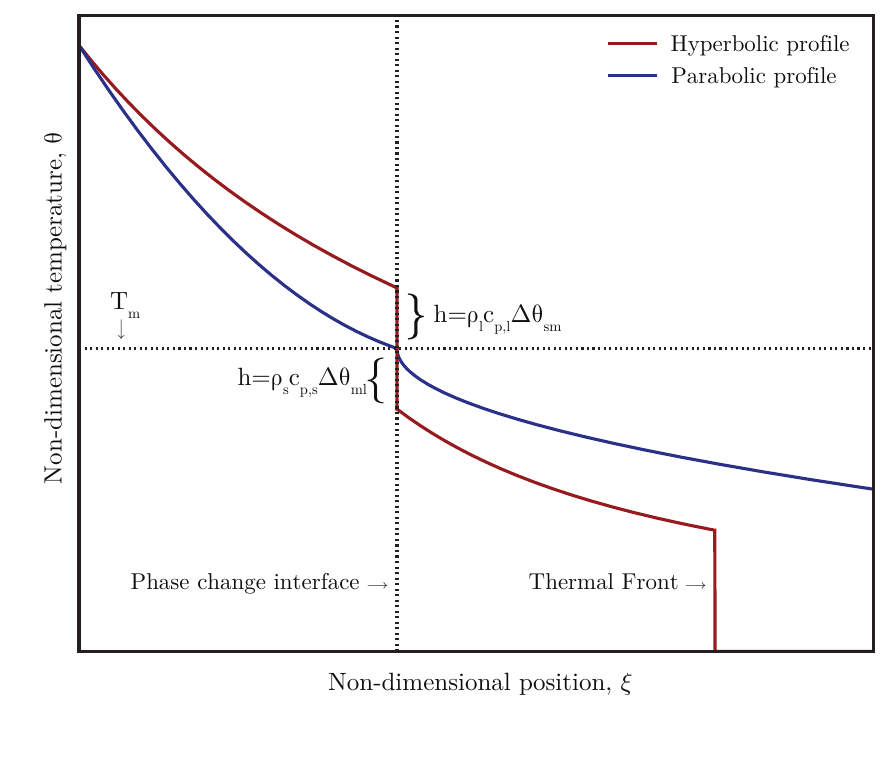}
    \caption{Comparison of Hyperbolic and Parabolic solutions to the classical 1D Stefan problem.}
\end{figure}

An additional factor arises when modelling systems at the micro and nanoscale: the thermophysical properties may not be considered constant and treated as bulk properties. Size-dependent latent heat and melting temperature depression models \cite{font2018variableproperties,font2019nonlocal} have been previously implemented in confined geometries (generally spherical), particularly in nanoscale melting problems where confinement alters the local interfacial thermodynamics. Including size-dependent physical properties and accounting for local, finite-size effects properly capture hyperbolic effects induced by premelting at the interface front, enabling a more physically-motivated numerical and analytic model.

The objective of this work is to develop a consistent analytic and numerical framework for the one-dimensional hyperbolic Stefan problem governed by the MCV equation with size-dependent thermo-physical system parameters in the limit of a small Stefan number (Ste $\ll$ 1). This assumption of a small Stefan number is motivated both by the perturbation series technique employed to obtain the solutions presented herein, and by physical relevance, as low Stefan numbers are commonly observed in phase-change systems. First, we derive perturbation approximation solutions using a reformulation in which time is expressed as a function of the solid-liquid interface position in both a parabolic and hyperbolic interface condition framework. Although the parabolic interface framework yields a simpler solution, it demonstrates an asymptotic divergence present in the early time limit ($t\to 0$). This is rectified by introducing an ``inner time'' solution, derived under a different time-scale, while using a matching method with a smoothing function to create a composite solution, for both temperature and the interface position, which is valid across the entire temporal domain (and ultimately the whole interface position domain). We also introduce an analysis of the numerical approach used in the verification simulations. Namely, a predictor-corrector MacCormack scheme, with numerically selective artificial viscosity to achieve total variation diminishing (TVD) behavior \cite{sweby1984high,harten1983high}. The TVD scheme suppresses spurious oscillations near thermal and wave fronts that arise in standard methods applied to hyperbolic moving boundary problems. The hyperbolic numerical model was verified in the parabolic limit (where the thermal relaxation time approaches zero) against diffusive simulations which can be implemented with a Forward-Time Centered-Space (FTCS) discretization, which avoids complicated matrix assembly that nonlinear mapping in the moving boundary tracking imposes on methods such as Crank–Nicolson.

The analytic approximations are then verified against the developed numerical method across a range of various Stefan numbers (Ste), thermal relaxation times ($\widetilde \tau$), and thermal diffusivities ($\alpha$). The combined framework clarifies the interplay between hyperbolic relaxation effects and classical diffusive behavior, identifies the regime of validity of the perturbation expansions, and provides a practical computational approach for non-Fourier phase change modeling. The contributions of the work are: (i) a matched asymptotic construction for the hyperbolic Stefan problem under a diffusive Stefan condition, (ii) incorporation of size-dependent interfacial thermodynamics within a non-diffusive Stefan condition, and (iii) an outline of a stabilized high-resolution numerical methodology suitable for wave-dominated phase change dynamics.

\section{Methodology}

\subsection{Formulation}
Consider a one-dimensional finite domain in the Cartesian coordinate, initially in the liquid form at its fusion temperature $T_{\mathrm{f}}$. At time $t = 0$, the temperature on the left side ($x=0$) is subjected to a cold surface that is constant at $T_{\mathrm{c}}$, where $T_{\mathrm{c}}<T_{\mathrm{f}}$. On the right side ($x = X$), the temperature is fixed to its initial temperature $T_{\mathrm{f}}$. The domain will start to freeze from the left to right; a solid-liquid interface or freezing front, $x_{\mathrm{i}}(t)$, will move over time in the same direction. For the purpose of mathematical simplification, the following assumptions are made: (i) there is no volumetric change during freezing and the mass densities of solid and liquid are assumed to be constant; (ii) the rest of thermophysical properties (e.g., $k$, $c_p$) are phase-dependent; and (iii) heat conduction is considered to be the predominant mode of heat transfer, particularly in a non-Fourier form, and the effect of natural convection is neglected.

The temperature profile in the solid phase is described by the Maxwell-Cattaneo-Vernotte (MCV) equation:
\begin{align}
    \frac{\partial^2 T}{\partial x^2}
    &=
    \frac{\tau}{\alpha}\frac{\partial^2 T}{\partial t^2}
    +
    \frac{1}{\alpha}\frac{\partial T}{\partial t},
    \qquad 0<x<x_{\mathrm{i}}(t),\quad t>0,
\end{align}
where $\tau$ is the thermal relaxation time [$\mathrm{s}$], which characterizes the finite delay between the onset of a temperature gradient and its corresponding heat flux, $\alpha$ is the thermal diffusivity $[\mathrm{m}^2\cdot\mathrm{s}^{-1}]$ calculated from $k/(\rho \cdot c_{p})$. $\rho$ and $c_p$ are the mass density $[\mathrm{kg}\cdot\mathrm{m}^{-3}]$ and specific heat $[\mathrm{J}\cdot\mathrm{kg}^{-1}\cdot\mathrm{K}^{-1}]$, respectively. It is noted that the addition of the second derivative with respect to time turns the heat conduction equation from parabolic (diffusive) to hyperbolic (non-diffusive). With the dynamic solid-liquid interface condition (the Stefan condition) expressed with a constant latent heat
\begin{equation}
    k \frac{\partial T}{\partial x} \bigg|_{x = x_{\mathrm{i}} (t)}
    = \rho L^* \frac{\mathrm{d} x_{\mathrm{i}} (t)}{\mathrm{d} t},
    ~0<x<X,~t>0.
\end{equation}
This formulation enables hyperbolic effects in the materials, but not at the phase change interface. This is an approximation for analytic tractability, although a non-diffusive framework for the Stefan condition will be introduced in Section 2.2.4. The system above in its standard form is subject to the following boundary and initial conditions (along with the Stefan condition)
\begin{align}
T(x=0,t)=T_c,\qquad T(x=X,t)=T_{f},\qquad T(x,t=0)=T_f
\end{align}
where $T_c$ [$\mathrm{K}$] is the temperature of the applied cold flux, and $T_f$ [$\mathrm{K}$] is the fusion temperature of the liquid.

The system solutions (temperature profile and interface profile) to the problem examined in this paper exist within this framework. That is, they are obtained when solving the MCV equation, subject to the boundary and initial conditions above, incorporating the size-dependent thermophysical values when needed in attempt to model additional non-Fourier effects into the phase change problem.

\subsection{Analytical modelling}
\subsubsection{Non-dimensionalization of the System}

The non-Fourier phase change problem can be non-dimensionalized by introducing the following variables:
\begin{equation}
    \mathrm{Ste} = \frac{c_p (T_{\mathrm{f}} - T_{c})}{L^{*}},~
    \widetilde t = \frac{\alpha t}{X^2} \times \mathrm{Ste},~
    \widetilde \tau = \frac{\alpha \tau}{X^2} \times \mathrm{Ste},~
    \widetilde T = \frac{T - T_c}{T_{\mathrm{f}} - T_c},~
    \widetilde x = \frac{x}{X},~
    \widetilde x_{\mathrm{i}} = \frac{x_{\mathrm{i}}}{X},
\end{equation}
where $\mathrm{Ste}$ is the Stefan number. $\widetilde t$, $\widetilde \tau$, $\widetilde T$, $\widetilde x$ and $\widetilde x_{\mathrm{i}}$ are the dimensionless time, relaxation time, temperature, $x$-coordinate, and interface locations, respectively. It is noted that the time $\widetilde t$ depends on the interface $\widetilde x_{\mathrm{i}}$, which has been shown to be more convenient than $\widetilde x_{\mathrm{i}} (\widetilde t)$ for an explicit form of the perturbation solutions \cite{brosa2019extended,mccue2008classical}. 

Rewriting the system in the non-dimensionalized variables defined in Eq. 4, the dimensionless system becomes
\begin{align}
    \frac{\partial^2 \widetilde T}{\partial \widetilde x^2}
    &=
    \widetilde\tau\,\mathrm{Ste}\,
    \frac{\partial^2 \widetilde T}{\partial \widetilde x_{\mathrm{i}}^2}
    \bigg(\frac{\mathrm{d}\widetilde t}{\mathrm{d}\widetilde x_{\mathrm{i}}}\bigg)^{-2}
    +
    \mathrm{Ste}\,
    \frac{\partial \widetilde T}{\partial \widetilde x_{\mathrm{i}}}
    \bigg(\frac{\mathrm{d}\widetilde t}{\mathrm{d}\widetilde x_{\mathrm{i}}}\bigg)^{-1},
    \qquad
    0<\widetilde x<\widetilde x_{\mathrm{i}},
\end{align}
subject to
\begin{align}
    \widetilde T(0,\widetilde x_{\mathrm{i}}) &= -1,
    &
    \widetilde T(\widetilde x_{\mathrm{i}},\widetilde x_{\mathrm{i}})&=0,\\
    \frac{\partial \widetilde T}{\partial \widetilde x}\bigg|_{\widetilde x=\widetilde x_{\mathrm{i}}}
    &=
    \bigg(\frac{\mathrm{d}\widetilde t}{\mathrm{d}\widetilde x_{\mathrm{i}}}\bigg)^{-1},
    &
    \widetilde t(\widetilde x_{\mathrm{i}}=0)&=0.
\end{align}
\subsubsection{Diffusive Stefan Condition Model}
The non-linearity of the moving boundary condition makes an exact analytical solution intractable, even when Fourier heat conduction is assumed at the interface. However, a key assumption of a small Stefan number is commonly made, as the contribution of latent heat is often much more than that of sensible heat during phase change. Assuming a small Stefan number, i.e., $\mathrm{Ste} \ll 1$, perturbation series solutions for both the temperature and time can be constructed:
\begin{align}
    \widetilde T &\sim \widetilde T_0 
    + \mathrm{Ste} \times \widetilde T_1 
    + \mathrm{Ste}^2 \times \widetilde T_2 +...\\
    \widetilde t &\sim \widetilde t_0 
    + \mathrm{Ste} \times \widetilde t_1
    + \mathrm{Ste}^2 \times \widetilde t_2 +...
\end{align}
When the approximate solutions (Eqs. 8 and 9) are substituted into the governing equation (Eq. 5), the problem at the leading order ($\mathcal{O}(1))$, reduces to:
\begin{align}
    \frac{\partial^2 \widetilde T_0}{\partial \widetilde x^2} &= 0,
    \qquad 0<\widetilde x<\widetilde x_{\mathrm{i}},
\end{align}
with boundary and initial conditions:
\begin{align}
    \widetilde T_0(0,\widetilde x_{\mathrm{i}})&=-1,
    &
    \widetilde T_0(\widetilde x_{\mathrm{i}},\widetilde x_{\mathrm{i}})&=0,\\
    \frac{\partial \widetilde T_0}{\partial \widetilde x}\bigg|_{\widetilde x=\widetilde x_{\mathrm{i}}}
    &=
    \bigg(\frac{\mathrm{d}\widetilde t_0}{\mathrm{d}\widetilde x_{\mathrm{i}}}\bigg)^{-1},
    &
    \widetilde t_0(0)&=0.
\end{align}
When solving for the leading order solutions
\begin{align}
    \widetilde T_0(\widetilde x,\widetilde x_{\mathrm{i}})
    &=
    -1+\frac{\widetilde x}{\widetilde x_{\mathrm{i}}},
    \qquad
    \widetilde t_0(\widetilde x_{\mathrm{i}})
    =
    \frac{\widetilde x_{\mathrm{i}}^{\,2}}{2},
\end{align}
which coincides with the classical one-phase Stefan leading-order solution. 

At the first order $(\mathcal{O}(Ste))$, the problem reduces to:
\begin{align}
    \frac{\partial^2 \widetilde T_1}{\partial \widetilde x^2}
    &=
    \widetilde\tau\,
    \frac{\partial^2 \widetilde T_0}{\partial \widetilde x_{\mathrm{i}}^2}
    \bigg(\frac{\mathrm{d}\widetilde t_0}{\mathrm{d}\widetilde x_{\mathrm{i}}}\bigg)^{-2}
    +
    \frac{\partial \widetilde T_0}{\partial \widetilde x_{\mathrm{i}}}
    \bigg(\frac{\mathrm{d}\widetilde t_0}{\mathrm{d}\widetilde x_{\mathrm{i}}}\bigg)^{-1},
    \qquad 0<\widetilde x<\widetilde x_{\mathrm{i}},
\end{align}
with boundary and initial conditions:
\begin{align}
    \widetilde T_1(0,\widetilde x_{\mathrm{i}})=0, 
    \widetilde T_1(\widetilde x_{\mathrm{i}},\widetilde x_{\mathrm{i}})=0,\\
    \frac{\mathrm{d}\widetilde t_0}{\mathrm{d}\widetilde x_{\mathrm{i}}}
    \frac{\partial \widetilde T_1}{\partial \widetilde x}\bigg|_{\widetilde x=\widetilde x_{\mathrm{i}}}
    +
    \frac{\mathrm{d}\widetilde t_1}{\mathrm{d}\widetilde x_{\mathrm{i}}}
    \frac{\partial \widetilde T_0}{\partial \widetilde x}\bigg|_{\widetilde x=\widetilde x_{\mathrm{i}}}
    =0,
    \qquad
    \widetilde t_1(0)=0.
\end{align}
Solving for the first order solutions
\begin{align}
    \widetilde T_1(\widetilde x,\widetilde x_{\mathrm{i}})
    &=
    \frac{\widetilde \tau\,\widetilde x^3}{3 \widetilde x_{\mathrm{i}}^{5}}
    +\frac{-2\widetilde \tau\,\widetilde x-\widetilde x^3}{6\widetilde x_{\mathrm{i}}^{3}}
    +\frac{\widetilde x}{6\widetilde x_{\mathrm{i}}},
\end{align}
\begin{align}
    \widetilde t_1(\widetilde x_{\mathrm{i}})
    =
    \frac{\widetilde x_{\mathrm{i}}^{\,2}}{6}
    -\frac{2}{3}\,\widetilde\tau\,\ln \widetilde x_{\mathrm{i}} + c.
\end{align}
Yielding the following perturbation solutions:
\begin{equation}
\widetilde{T}(\tilde x,\tilde x_\mathrm{i})=-1+\frac{\tilde x}{\tilde x_\mathrm{i}} +\mathrm{Ste}\left[ \frac{\widetilde \tau\,\widetilde x^3}{3 \widetilde x_{\mathrm{i}}^{5}}
    +\frac{-2\widetilde \tau\,\widetilde x-\widetilde x^3}{6\widetilde x_{\mathrm{i}}^{3}}
    +\frac{\widetilde x}{6\widetilde x_{\mathrm{i}}}\right],
\end{equation}
\begin{equation}
    \tilde t(\tilde x,\tilde x_\mathrm{i})=\frac{\widetilde x_{\mathrm{i}}^{\,2}}{2}+\mathrm{Ste}\left[  \frac{\widetilde x_{\mathrm{i}}^{\,2}}{6}
    -\frac{2}{3}\,\widetilde\tau\,\ln \widetilde x_{\mathrm{i}} + c\right].
\end{equation}
The interfacial correction $\widetilde t_1$ obtained from the $\mathcal{O}(\mathrm{Ste})$ Stefan condition contains a logarithmic term which diverges as $\widetilde x_{\mathrm{i}}\to 0^+$, and therefore cannot satisfy the physical limit of the initial condition ($\lim_{\widetilde x_{\mathrm{i}}\to0}\tilde t(x_{\mathrm{i}})=0 $). Two approaches to rectify this asymptotic divergence are explored. Firstly, the introduction of an inner initial layer, valid in a rescaled temporal domain, which can properly capture the early time dynamics of the moving interface. Secondly, the hyperbolic effects at the phase change interface are included in the model in the way of size-dependent thermophysical values. 

\subsubsection{Inner Solution to Diffusive Stefan Condition Model}
The perturbation series solution above fails to capture the early-time ($\mathcal{O}$(Ste)) dynamics of the phase change boundary due to the omission of the hyperbolic relaxation term in the limit of Ste $\to0$, and consequently does not capture the initial condition. The first proposed fix to this divergence present at early-times is the derivation of an inner solution, valid on small time scales, that can be coupled with the outer solution to preserve the initial condition. This rescaling of the temporal variable should be in such a way that the hyperbolic term of the governing equation is included in the first order solution. To this end, the following rescaled ``fast'' time is introduced:

\begin{align}
\widetilde{t}=\sqrt{\mathrm{Ste}}\,\phi,
\qquad
\frac{\partial}{\partial \widetilde{t}}
=\frac{1}{\sqrt{\mathrm{Ste}}}\frac{\partial}{\partial \phi},
\qquad
\frac{\partial^2}{\partial \widetilde{t}^2}
=\frac{1}{\mathrm{Ste}}\frac{\partial^2}{\partial \phi^2}.
\label{eq:inner_scaling}
\end{align}
Substitution into the dimensionless MCV equation yields
\begin{align}
\frac{\partial^2 \widetilde{T}}{\partial \widetilde{x}^2}
&=
\widetilde{\tau}\,\frac{\partial^2 \widetilde{T}}{\partial \phi^2}
+\sqrt{\mathrm{Ste}}\,
\frac{\partial \widetilde{T}}{\partial \phi},
\qquad 0<\widetilde{x}<\widetilde{x}_{\mathrm{i}}(\phi),
\label{eq:inner_pde}
\end{align}
with the same boundary conditions and initial conditions:
\begin{align}
\widetilde{T}(0,\phi)&=-1,
&
\widetilde{T}(\widetilde{x}_{\mathrm{i}},\phi)&=0,
\label{eq:inner_bc}
\end{align}
\begin{align}
\frac{\partial \widetilde{T}}{\partial \widetilde{x}}
\bigg|_{\widetilde{x}=\widetilde{x}_{\mathrm{i}}(\phi)}
&=
\frac{\mathrm{d}\tilde{x_\mathrm{i}}}{\mathrm{d}\widetilde t}
=
\frac{1}{\sqrt{\mathrm{Ste}}}
\frac{\mathrm{d}\tilde x_{\mathrm{i}}}{\mathrm{d}\phi}
,\qquad \phi(0)=0.
\label{eq:inner_stefan}
\end{align}

We seek inner expansions in powers of $\sqrt{\mathrm{Ste}}$:
\begin{align}
\widetilde{T}
&\sim
\widetilde{T}_{0,\mathrm{in}}
+\sqrt{\mathrm{Ste}}\,
\widetilde{T}_{1,\mathrm{in}}
+\mathrm{Ste} \widetilde{T}_{2,\mathrm{in}}...,
\label{eq:inner_T_exp}
\end{align}
The temperature profile coefficients, $\widetilde T_{\mathrm{i},\mathrm{in}}$, are obtained order by order, each admitting a closed-form modal solution. The interface profile, however, will prove to require a numerical solver, and thus a term-by-term splitting provides no analytic benefit.
Substituting Eq.~\eqref{eq:inner_T_exp} into Eq.~\eqref{eq:inner_pde}
and collecting $\mathcal{O}(1)$ terms in the limit of Ste $\to 0$ yields the leading-order inner equation
\begin{align}
\frac{\partial^2 \widetilde{T}_{0,\mathrm{in}}}{\partial \widetilde{x}^2}
=
\widetilde{\tau}\,
\frac{\partial^2 \widetilde{T}_{0,\mathrm{in}}}{\partial \phi^2},
\qquad 0<\widetilde{x}<\widetilde{x}_{\mathrm{i}}(\phi),
\label{eq:inner_leading_wave}
\end{align}
with boundary and initial conditions
\begin{equation}
\begin{aligned}
\widetilde{T}_{0,\mathrm{in}}(0,\phi) &= -1,
&\qquad
\widetilde{T}_{0,\mathrm{in}}(\widetilde{x}_{\mathrm{i}},\phi) &= 0,\\
\frac{\partial \widetilde T_0}{\partial \tilde x}\bigg|_{\tilde x=\tilde x_\mathrm{i}}
&=
\frac{1}{\sqrt{\mathrm{Ste}}}\frac{d\tilde x_\mathrm{i}}{d\tilde \phi},
&\qquad
\widetilde T_{0,in}(\tilde x_\mathrm{i},0)&=0.
\end{aligned}
\label{eq:inner_leading_bc}
\end{equation}

Eq.~\eqref{eq:inner_leading_wave} is a wave equation in the inner time
$\phi$ with wave speed $c=1/\sqrt{\widetilde{\tau}}$.
We homogenize the Dirichlet boundary conditions by writing
\begin{align}
\widetilde{T}_{0,\mathrm{in}}(\widetilde{x},\phi)
&=
-1+\frac{\widetilde{x}}{\widetilde{x}_{\mathrm{i}}}
+v(\widetilde{x},\phi),
\qquad
v(0,\phi)=v(\widetilde{x}_{\mathrm{i}},\phi)=0,
\label{eq:inner_homogenize}
\end{align}
when using a sinusoidal eigenfunction expansion, the homogeneous component becomes
\begin{align}
v(\widetilde{x},\phi)
&=
\sum_{n=1}^{\infty}
\left[
A_n \cos\!\left(\frac{n\pi c\,\phi}{\widetilde{x}_{\mathrm{i}}}\right)
+
B_n \sin\!\left(\frac{n\pi c\,\phi}{\widetilde{x}_{\mathrm{i}}}\right)
\right]
\sin\!\left(\frac{n\pi \widetilde{x}}{\widetilde{x}_{\mathrm{i}}}\right).
\label{eq:inner_v_series}
\end{align}
If the initial temperature in the solid is taken to be $\widetilde{T}(\widetilde{x},0)=0$ (hence
$v(\widetilde{x},0)=1-\widetilde{x}/\widetilde{x}_{\mathrm{i}}$), then the cosine coefficients satisfy
\begin{align}
A_n=\frac{2}{\pi n}, \qquad n=1,2,\dots
\label{eq:inner_An}
\end{align}
and the leading-order inner temperature may be written as
\begin{align}
\widetilde{T}_{0,\mathrm{in}}(\widetilde{x},\phi)
&=
-1+\frac{\widetilde{x}}{\widetilde{x}_{\mathrm{i}}}
+
\sum_{n=1}^{\infty}
\left[
\frac{2}{\pi n}\cos\!\left(\frac{n\pi c\,\phi}{\widetilde{x}_{\mathrm{i}}}\right)
+
B_n \sin\!\left(\frac{n\pi c\,\phi}{\widetilde{x}_{\mathrm{i}}}\right)
\right]
\sin\!\left(\frac{n\pi \widetilde{x}}{\widetilde{x}_{\mathrm{i}}}\right).
\label{eq:inner_T0_final}
\end{align}
In contrast to the previous formulations, a closed solution for the interface position in time is not possible in either the leading or first order. Applying the same substitution used in Section 2.2.2 leads to degeneracy in the ``fast'' time domain. The interface position must thus be solved for implicitly.

The leading-order outer solution obtained in Section 2.2.2 has the form
$\widetilde{T}_{0,\mathrm{out}}=-1+\widetilde{x}/\widetilde{x}_{\mathrm{i}}$.
Therefore, a unique leading order solution, $\widetilde T_{0,\mathrm{in}}$, may be obtained with the Van Dyke matching condition \cite{VanDyke1975}:
\begin{align}
\lim_{\phi\to\infty}\widetilde{T}_{0,\mathrm{in}}(\widetilde{x},\phi)
=
\lim_{\widetilde{t}\to 0^+}\widetilde{T}_{0,\mathrm{out}}(\widetilde{x},\widetilde{t}),
\label{eq:inner_matching}
\end{align}
requires that the oscillatory contributions in Eq.~\eqref{eq:inner_T0_final}
do not contribute to the common part. This implies:
\begin{align}
B_n=0,\qquad n=1,2,\dots
\label{eq:inner_Bn_zero}
\end{align}
at the leading order, so that the matched inner temperature relaxes to the
outer profile as $\phi\to\infty$.

In the same limit of Ste $\to0$, the first order inner equation becomes a forced-wave equation
\begin{equation}
    \frac{\partial^2 \widetilde{T}_{1,\mathrm{in}}}{\partial \widetilde{x}^2}
    =
    \widetilde{\tau}\,
    \frac{\partial^2 \widetilde{T}_{1,\mathrm{in}}}{\partial \phi^2}
    +
    \frac{\partial \widetilde{T}_{0,\mathrm{in}}}{\partial \phi},
    \qquad 0<\widetilde{x}<\widetilde{x}_{\mathrm{i}},
\end{equation}
with homogeneous boundary conditions to preserve the boundary conditions in the perturbation series solution:
\begin{align}
\widetilde{T}_{1,\mathrm{in}}(0,\phi)&=0,
&
\widetilde{T}_{1,\mathrm{in}}(\widetilde{x}_{\mathrm{i}},\phi)&=0.
\end{align}
and with initial conditions
\[
\frac{\partial\left(\widetilde T_0+\sqrt{\mathrm{Ste}} \widetilde T_1\right)}{\partial \tilde x}\bigg|_{\tilde x=\tilde x_\mathrm{i}}
=\frac{1}{\sqrt{\mathrm{Ste}}}
\frac{d\tilde x_\mathrm{i}}{d\tilde \phi},
\qquad
\widetilde T_{0,\mathrm{in}}(\tilde x_\mathrm{i},0)=0.
\]
Differentiating the leading order solution gives the source term:
\begin{equation}
    \frac{\partial \widetilde{T}_{0,\mathrm{in}}}{\partial \phi}
    =
    -\sum_{n=1}^{\infty}
    \frac{2c}{\widetilde{x}_{\mathrm{i}}}
    \sin\!\left(\frac{n\pi c\,\phi}{\widetilde{x}_{\mathrm{i}}}\right)
    \sin\!\left(\frac{n\pi \widetilde{x}}{\widetilde{x}_{\mathrm{i}}}\right).
\end{equation}
This is a linear combination of the spatial eigenfunctions, $\mathrm{sin}(n\pi \tilde x/\tilde x_{\mathrm{i}})$. The boundary conditions of $\widetilde T_{1,in}$  are homogeneous, enabling a similar eigenfunction basis expansion:
\[
\widetilde{T}_{1,\mathrm{in}}=\sum_{n=1}^{\infty} b_n(\phi)
\sin(n\pi \widetilde{x}/\widetilde{x}_{\mathrm{i}})
\]
which yields an ODE for each modal coefficient, $b_n$:
\[
b_n''+\omega ^2_n b_n=\frac{2c}{\widetilde \tau \widetilde x_\mathrm{i}}\mathrm{sin}(\omega_n \phi),\qquad\omega_n=\frac{n\pi c}{\tilde x_\mathrm{i}}
\]
with a solution, under $b_n(0)=b_n'(0)=0$ (constant initial temperature profile),
\begin{equation}
    b_n(\phi)
    =
    \frac{1}{\pi n\,\widetilde{\tau}}
    \left[
    \frac{\sin\!\left(\omega_n \phi\right)}{\omega_n}
    -
    \phi\,\cos\!\left(\omega_n \phi\right)
    \right].
\end{equation}
The first-order inner temperature profile coefficient is therefore
\begin{equation}
    \widetilde{T}_{1,\mathrm{in}}(\widetilde{x},\phi)
    =
    \sum_{n=1}^{\infty}
    \frac{1}{\pi n\,\widetilde{\tau}}
    \left[
    \frac{\sin\!\left(\omega_n \phi\right)}{\omega_n}
    -
    \phi\,\cos\!\left(\omega_n \phi\right)
    \right]
    \sin\!\left(\frac{n\pi \widetilde{x}}{\widetilde{x}_{\mathrm{i}}}\right)...
\end{equation}
The perturbation solutions of the inner problem are
\begin{equation}
\widetilde{T}(\widetilde{x},\phi)
=
-1
+\frac{\widetilde{x}}{\widetilde{x}_{\mathrm{i}}}
+
\sum_{n=1}^{\infty}
\left[
A_n(\phi)
+
\sqrt{\mathrm{Ste}}\,C_n(\phi)
\right]
\sin\!\left(
\frac{n\pi\widetilde{x}}
{\widetilde{x}_{\mathrm{i}}}
\right),
\end{equation}

\begin{equation}
\frac{d\widetilde{x}_{\mathrm{i}}}{d\phi}
=
\sqrt{\mathrm{Ste}}\left[\frac{1}{\widetilde{x}_{\mathrm{i}}}
+
\frac{\pi}{\widetilde{x}_{\mathrm{i}}}
\sum_{n=1}^{\infty}
n(-1)^n
\left[
A_n(\phi)
+
\sqrt{\mathrm{Ste}}\,C_n(\phi)
\right]\right].
\end{equation}
where
\begin{align}
A_n(\phi)
&=
\frac{2}{\pi n}
\cos\!\left(
\frac{n\pi c\,\phi}
{\widetilde{x}_{\mathrm{i}}}
\right),
\\[6pt]
C_n(\phi)
&=
\frac{1}{\pi n\,\widetilde{\tau}}
\left[
\frac{\sin\!\left(\omega_n\phi\right)}
{\omega_n}
-
\phi\cos\!\left(\omega_n\phi\right)
\right].
\end{align}

Composite solution profiles are not obtainable with the traditional Van Dyke matching rule due to the asymptotic divergence of the outer solution \cite{VanDyke1975}. However, a smoothing function with the following properties may be used to obtain a solution valid across the entire interface domain ($\widetilde x_i\in[0,1]$). 
\begin{align}
t_{\mathrm{comp}}(\widetilde{x}_\mathrm{i})
&=
u(\widetilde{x}_\mathrm{i})\,t_{\mathrm{in}}(\widetilde{x}_\mathrm{i})
+
\bigl[1-u(\widetilde{x}_\mathrm{i})\bigr]\,
t_{\mathrm{out}}(\widetilde{x}_i),
\label{eq:blended_comp}
\end{align}
where
\begin{align}
0 \le u(\widetilde{x}_\mathrm{i}) \le 1, \qquad
\lim_{\widetilde{x}_\mathrm{i} \to 0} u(\widetilde{x}_\mathrm{i}) = 1, \qquad
\lim_{\widetilde{x}_\mathrm{i} \to 1} u(\widetilde{x}_\mathrm{i}) = 0.
\end{align}
The initial condition of the composite equation is captured by the inner solution, hence the integration constant present in Eq. 20 can be dropped altogether.
\subsubsection{Non-Diffusive Stefan Condition Model}
The second approach examined in this paper to rectify the asymptotic divergence takes into account non-Fourier effects at the phase change interface, which can be realized with the introduction of size-dependent thermophysical values (Latent heat of phase change and the melting temperature). If the melting temperature is assumed to remain constant in the analytic model ($T (x = x_{\mathrm{i}} (t), t) = T_\mathrm{f}$), the Stefan condition becomes:
\begin{equation}
    k \frac{\partial T}{\partial x} \bigg|_{x = x_{\mathrm{i}} (t)}
    = \rho L (t) \frac{\mathrm{d} x_{\mathrm{i}} (t)}{\mathrm{d} t},
    ~0<x<X,~t>0
\end{equation}
where $L(t)$ can be a function of any shape that reflects the effective latent heat, written generically as:
\begin{equation}
    L(t) = L^{*}\cdot f({x_{\mathrm{i}}(t)}),
\end{equation}
Incorporating the general form presented in Lai et al. \cite{lai1996}, the Stefan condition can be written as
\begin{equation}
    k \frac{\partial T}{\partial x} \bigg|_{x = x_{\mathrm{i}} (t)}
    = \rho L^* \frac{\mathrm{d} x_{\mathrm{i}} (t)}{\mathrm{d} t} 
    +
    \rho \bigg[ 1 - \frac{1}{x_{\mathrm{i}} (t)} \bigg] \frac{\mathrm{d} x_{\mathrm{i}} (t)}{\mathrm{d} t}
\end{equation}
or expressed in the dimensionless framework
\[
\frac{\partial \widetilde T}{\partial \widetilde x} \bigg|_{\widetilde x = \widetilde x_{\mathrm{i}}}  
    = 
    \bigg(
    \frac{\mathrm{d} \widetilde t}{\mathrm{d} \widetilde x_{\mathrm{i}}}
    \bigg)^{-1}
    + \bigg[ 1 - \frac{1}{X \widetilde x_{\mathrm{i}}} \bigg] \frac{1}{L^*}
    \bigg(
    \frac{\mathrm{d} \widetilde t}{\mathrm{d} \widetilde x_{\mathrm{i}}}
    \bigg)^{-1}
\]
Thus, the dimensionless problem is written as 
\begin{align}
    \frac{\partial^2 \widetilde T}{\partial \widetilde x^2} &= 
    \widetilde \tau \mathrm{Ste}
    \frac{\partial^2 \widetilde T}{\partial \widetilde x_{\mathrm{i}}^2}
    \bigg(
    \frac{\mathrm{d} \widetilde t}{\mathrm{d} \widetilde x_{\mathrm{i}}}
    \bigg)^{-2}
    +
    \mathrm{Ste} \frac{\partial \widetilde T}{\partial \widetilde x_{\mathrm{i}}}
    \bigg(
    \frac{\mathrm{d} \widetilde t}{\mathrm{d} \widetilde x_{\mathrm{i}}}
    \bigg)^{-1}
    ,~0<\widetilde x<\widetilde x_{\mathrm{i}},
\end{align}
subject to
\begin{align}
    \widetilde T (\widetilde x = 0, \widetilde x_{\mathrm{i}}) &= -1, &
    \widetilde T (\widetilde x = \widetilde x_{\mathrm{i}}, \widetilde x_{\mathrm{i}}) &= 0, \\
    \frac{\partial \widetilde T}{\partial \widetilde x} \bigg|_{\widetilde x = \widetilde x_{\mathrm{i}}}  
    &= 
    \bigg(
    \frac{\mathrm{d} \widetilde t}{\mathrm{d} \widetilde x_{\mathrm{i}}}
    \bigg)^{-1}
    + \bigg[ 1 - \frac{1}{X \widetilde x_{\mathrm{i}}} \bigg] \frac{1}{L^*}
    \bigg(
    \frac{\mathrm{d} \widetilde t}{\mathrm{d} \widetilde x_{\mathrm{i}}}
    \bigg)^{-1}
    , &
    \widetilde t (\widetilde x_{\mathrm{i}} = 0) &= 0
\end{align}
Assuming a small Stefan number, i.e., the limit of $Ste\to 0$, we can construct perturbation series solutions for both temperature and time, similar to in Section 2.2.2:
\begin{align}
    \widetilde T &\sim \widetilde T_0
    +\mathrm{Ste}\,\widetilde T_1
    +\mathrm{Ste}^2\,\widetilde T_2+\cdots,\\
    \widetilde t &\sim \widetilde t_0
    +\mathrm{Ste}\,\widetilde t_1
    +\mathrm{Ste}^2\,\widetilde t_2+\cdots.
\end{align}

At the leading order, the problem becomes:
\begin{align}
    \frac{\partial^2 \widetilde T_0}{\partial \widetilde x^2} &= 
    0
    ,~0<\widetilde x<\widetilde x_\mathrm{i}
\end{align}
subject to
\begin{align}
    \widetilde T_0 (\widetilde x = 0, \widetilde x_\mathrm{i}) &= -1, &
    \widetilde T_0 (\widetilde x = \widetilde x_\mathrm{i}, \widetilde x_\mathrm{i}) &= 0, \\
    \frac{\partial \widetilde T_0}{\partial \widetilde x} \bigg|_{\widetilde x = \widetilde x_{\mathrm{i}}}  
    &= 
    \bigg(
    \frac{\mathrm{d} \widetilde t_0}{\mathrm{d} \widetilde x_{\mathrm{i}}}
    \bigg)^{-1}
    + \bigg[ 1 - \frac{1}{X \widetilde x_{\mathrm{i}}} \bigg] \frac{1}{L^*}
    \bigg(
    \frac{\mathrm{d} \widetilde t_0}{\mathrm{d} \widetilde x_{\mathrm{i}}}
    \bigg)^{-1}
    , &
    \widetilde t_0 (\widetilde x_\mathrm{i} = 0) &= 0.
\end{align}
The leading-order solutions of solid phase are thus the same in temperature as the diffusive interface condition formulation, differing only in the interface profile from the size-dependent Stefan condition:
\begin{equation}
    \widetilde T_0 =
    -1 + \frac{\widetilde x}{\widetilde x_\mathrm{i}}, ~
    \widetilde t_0 = 
    \frac{(1+L^{*}) \widetilde x_\mathrm{i}^2}{2L^{*}} 
    - 
    \frac{\widetilde x_\mathrm{i}}{XL^{*}}
    .
\end{equation}

For the first-order term, the problem becomes:
\begin{align}
    \frac{\partial^2 \widetilde T_1}{\partial \widetilde x^2} &= 
    \widetilde \tau 
    \frac{\partial^2 \widetilde T_0}{\partial \widetilde x_\mathrm{i}^2}
    \bigg(
    \frac{\mathrm{d} \widetilde t_0}{\mathrm{d} \widetilde x_\mathrm{i}}
    \bigg)^{-2}
    +
    \frac{\partial \widetilde T_0}{\partial \widetilde x_\mathrm{i}}
    \bigg(
    \frac{\mathrm{d} \widetilde t_0}{\mathrm{d} \widetilde x_\mathrm{i}}
    \bigg)^{-1}
    ,~0<\widetilde x<\widetilde x_\mathrm{i},
\end{align}
subject to
\begin{align}
    \widetilde T_1 (\widetilde x = 0, \widetilde x_\mathrm{i}) &= 0, &
    \widetilde T_1 (\widetilde x = \widetilde x_\mathrm{i}, \widetilde x_\mathrm{i}) &= 0, \\
    \frac{\mathrm{d} \widetilde t_0}{\mathrm{d} \widetilde x_\mathrm{i}}
    \frac{\partial \widetilde T_1}{\partial \widetilde x} \bigg|_{\widetilde x = \widetilde x_\mathrm{i}} +
    \frac{\mathrm{d} \widetilde t_1}{\mathrm{d} \widetilde x_\mathrm{i}}
    \frac{\partial \widetilde T_0}{\partial \widetilde x} \bigg|_{\widetilde x = \widetilde x_\mathrm{i}}
    &= 0, &
    \widetilde t_1 (\widetilde x_\mathrm{i} = 0) &= 0.
\end{align}
The first-order solutions of solid phase temperature and interface are:
\begin{align}
    \widetilde T_1 
    &=
    \frac{L^{*} X
    (\widetilde x \widetilde x_{\mathrm{i}}^2 - 
    \widetilde x ^3)
    [
    \widetilde x_{\mathrm{i}} (X \widetilde x_{\mathrm{i}} - 1)
    +
    L^{*} X
    (\widetilde x_{\mathrm{i}}^2 - 2 \widetilde \tau)
    ]
    }{6 \widetilde x_{\mathrm{i}}^3
    [
    (1+L^{*}) X \widetilde x_{\mathrm{i}} - 1
    ]^2
    },\\\\
    \widetilde t_1 
    &=
    \frac{ \widetilde x_{\mathrm{i}}^2}{6}
    -
    \frac{2 L^* \widetilde \tau \ln 
    [|1 - (1+ L^{*})X \widetilde x_{\mathrm{i}}|]}{3 (1+L^{*})
    }.
\end{align}
In the limit of large $L^*$ (equivalently, $\mathrm{Ste}\to0$), this simplifies to
\[
    \frac{ \widetilde x_{\mathrm{i}}^2}{6}
    -
    \frac{2 L^* \widetilde \tau \ln 
    [|1 - (1+ L^{*})X \widetilde x_{\mathrm{i}}|]}{3 (1+L^{*})
    } \approx \frac{ \widetilde x_{\mathrm{i}}^2}{6}
    -
    \frac{2 L^* \widetilde \tau \ln 
    [1 + (1+ L^{*})X \widetilde x_{\mathrm{i}}]}{3 (1+L^{*})
    }, \qquad \forall\widetilde x_i > 0.
\]
Yielding the following perturbation solutions:
\begin{equation}
    \widetilde T(\tilde x,\tilde x_\mathrm{i})=-1+\frac{\tilde x}{\tilde x_\mathrm{i}} +\mathrm{Ste}\left[ \frac{L^{*} X
    (\widetilde x \widetilde x_{\mathrm{i}}^2 - 
    \widetilde x ^3)
    [
    \widetilde x_{\mathrm{i}} (X \widetilde x_{\mathrm{i}} - 1)
    +
    L^{*} X
    (\widetilde x_{\mathrm{i}}^2 - 2 \widetilde \tau)
    ]
    }{6 \widetilde x_{\mathrm{i}}^3
    [
    (1+L^{*}) X \widetilde x_{\mathrm{i}} - 1
    ]^2
    }\right]...
\end{equation}
\begin{equation}
    \tilde t(\tilde x,\tilde x_\mathrm{i})=\frac{(1+L^{*}) \widetilde x_\mathrm{i}^2}{2L^{*}} 
    - 
    \frac{\widetilde x_\mathrm{i}}{XL^{*}} +\mathrm{Ste}\left[\frac{ \widetilde x_{\mathrm{i}}^2}{6}
    -
    \frac{2 L^* \widetilde \tau \ln 
    [1 + (1+ L^{*})X \widetilde x_{\mathrm{i}}]}{3 (1+L^{*})
    }\right]...
\end{equation}
which are well defined over the interface domain ($x_i \in[0,1]$). Note that the early interface divergence of the interface series solution (Eq.~64) predicts negative time. This arises from the approximate nature of the perturbation solution construction; the leading order solution for the temperature profile encodes the finite onset delay of the hyperbolic effects in the way of an initial negative time divergence. This acts as a temporal shift to model the wave effects, which increases the models accuracy, though introducing non-physical results at small time values, as will be seen in the discussion.

\subsection{Numerical modelling}
The following numerical method builds upon the second-order explicit predictor-corrector MacCormack, modified to suppress spurious oscillations present in the original framework near thermal and melting wave fronts \cite{maccormack1982numerical}. This scheme was selected for its numerical robustness in wave-fronts and computational tractability, as it avoids Jacobian evaluations required by most implicit schemes, making it attractive for nonlinear problems \cite{Tannehill1997}.

In this study, we consider the one-dimensional hyperbolic Stefan problem outlined in the analytic formulation, implemented with the same dimensionless variables for consistency.
 The governing equations (in dimensional form) consist of the energy equation and the hyperbolic flux equation:
\begin{align}
\frac{\partial H}{\partial t} + \frac{\partial Q}{\partial x} &= 0, \\
\frac{\partial Q}{\partial t} + \frac{k}{\tau} \frac{\partial T}{\partial x} &= -\frac{2Q}{\tau},
\end{align}
where $H$ [$\mathrm{J}$] is the enthalpy, $Q$ [$\mathrm{W}\cdot \mathrm{m}^{-2}$] is the heat flux, $T$ [$\mathrm{K}$] is the temperature, $k$ is the thermal conductivity [$\mathrm{W}\cdot\mathrm{m}^{-1}\cdot\mathrm{K}^{-1}$], and $\tau$ is the thermal relaxation time[$\mathrm{s}$]. The "Enthalpy method" is used in numerical solutions as it allows for numerical interface tracking without explicitly enforcing the Stefan condition at each time step, thereby simplifying implementation \cite{voller1981accurate}.  To ensure numerical robustness and improve the performance of the enthalpy-based formulation, a smoothing parameter,
$\epsilon$, is incorporated and defined as
\begin{align} 
\epsilon =: T_m^+-T_m^-, \qquad
\end{align}
and chosen as a function of the spatial step
\[
\epsilon=f(\Delta x).
\]
The function $f(\Delta x)$ reflects the dependence of $\epsilon$ on the spatial discretization. In this work, linear interpolation is used to resolve the enthalpy-temperature map precisely to obtain a weak-form solution. The general enthalpy profile is:
\begin{align}
H(T _f)=
\int_{0}^{T_f} \left[ \psi(T,\epsilon) + \delta(T- T_m,\epsilon) \cdot L^* \right] \, dT,
\end{align}
where $\psi$ models the sensible heat contribution and $\delta(T - T_m)$ models the latent heat contribution, resembling a Delta-Dirac function. Given the governing equations are hyperbolic, a predictor-corrector MacCormack method is used. The system is written in conservative form as:
\begin{align}
U &= 
\begin{bmatrix}
H \\
Q
\end{bmatrix},\qquad
F = 
\begin{bmatrix}
Q \\
T
\end{bmatrix},
\end{align}
The predictor step (forward difference) is:
\begin{align}
U^{n+1,\mathrm{forward}}_i = U_i^n - \frac{\Delta t}{\Delta x} B\left(F_{i+1}^n - F_i^n\right) + \Delta t \cdot S(U_i^n),
\end{align}
where S is the source term that models flux decay due to thermal relaxation effects, and B is the Jacobian matrix:
\begin{align}
S(U) &= \begin{bmatrix} 0 \\ -\dfrac{2Q}{\tau} \end{bmatrix}, \qquad
B = \begin{bmatrix} 1 & 0 \\ 0 & \dfrac{k}{\tau} \end{bmatrix}.
\end{align}

The corrector step (backward difference) is:
\begin{align}
U^{n+1,\mathrm{backward}}_i = U_i^n - \frac{\Delta t}{\Delta x}B \left(F_{i}^{n+1} - F_{i-1}^{n+1}\right) + \Delta t \cdot S(U_i^{n+1}).
\end{align}
The final solution is computed as an average of the predictor and corrector steps:
\begin{align}
U_i^{n+1} &= \frac{1}{2} \left( U_i^n + U_i^{n+1, \text{forward}} - \frac{\Delta t}{\Delta x} B \left( F_i^{n+1} - F_{i-1}^{n+1} \right) + \Delta t \cdot S\left( U_i^{n+1} \right) \right).
\end{align}

This scheme is second-order accurate in  space and time with truncation errors of $\mathcal{O}(\Delta x^2)$ and $\mathcal{O}(\Delta t^2)$. However, the MacCormack method is prone to spurious oscillations reminiscent of the Gibbs phenomenon, a result of its low numerical dissipation. To mitigate these inaccuracies, a selective artificial viscosity is introduced and approximated as \cite{davis1984tvd}:
 $$\frac{\partial}{\partial x} \left( \mu(x) \frac{\partial U}{\partial x} \right) \approx D_{i+1/2} - D_{i-1/2},$$
where:
\begin{align}
D_{i+1/2} &= \left(G_i^+ + G_{i+1}^-\right) \Delta U_i^+, \\
D_{i-1/2} &= \left(G_{i-1}^+ + G_i^-\right) \Delta U_i^-,\\
\Delta U_i^+ &= U_{i+1} - U_i,\\
\Delta U_i^- &= U_i - U_{i-1}.
\end{align}
where the superscripts +, -, refer to the front or back of each cell, $i$. The effective numerical viscosity coefficients locally calculated with flux limiters:
$$G_i^{\pm} = \frac{C(\nu_i)}{2} \left(1 - \theta(r_i^{\pm})\right),$$
where $\theta$ represents a flux limiter function, $C(\nu_i)$ is a local Courant number function, and $r^{\pm}$ represents the forward and backward cell gradient ratios.

TVD can be achieved through the appropriate selection of flux limiter functions, a process effectively guided by a Sweby diagram \cite{sweby1984high}. Incorporating the artificial viscosity into the MacCormack scheme yields the following equation (MacCormack-TVD):
$$U_i^{n+1} = \frac{1}{2} \left(U_i^n + U_i^{n+1,\mathrm{forward}} - \frac{\Delta t}{\Delta x}B \left(F_{i}^{n+1} - F_{i-1}^{n+1}\right) + \Delta t \cdot S(U_i^{n+1}) + \lambda \left(D_{i+1/2} - D_{i-1/2}\right) \right),$$
where $\lambda$ is a control parameter for the artificial viscosity.

Fourier analysis of the discretized system yields restrictions on the Courant number and the viscous control factor, $\lambda$. The stability limits depend on the diffusivity of the selected flux limiter function. For example, the Minmod function:
\[
\phi_{\mathrm{minmod}}(r)= \mathrm{max}[0,\mathrm{min}(1,r)], \qquad \lim_{r \to \infty}\phi_{\mathrm{minmod}} (r)=1,
\]
which is among the most diffusive TVD limiters, introduces high numerical diffusion which suppresses sharp wave-fronts, allowing for stable operation at larger Courant numbers (CFL $\le$ 1). Given that the Jacobian matrix is invariant in time and space, a global Courant number can be used for computational efficiency. 

Up to this point, the interface condition has been treated within a diffusive framework. To incorporate hyperbolic effects and size-dependent values for the latent heat and melting temperature, we model the molten surface sub-layer as being in thermodynamic equilibrium with the underlying solid core. The differential volume of the molten surface sub-layer can be represented by:
\begin{equation}
    \delta V = l \cdot w \cdot \left(x_t - \gamma \right),
\end{equation}
where $l$ [$\mathrm{m}$] is the material slab thickness, $w$ [$\mathrm{m}$] is the width, $x_t$ [$\mathrm{m}$] is the total slab length, and $\gamma$ [$\mathrm{m}$] is the molten sub-layer length. Assuming the latent heat of the core to be independent of temperature, the differential volume can be combined with the core volume to yield the following size-dependent functions \cite{lai1996}:
\begin{equation}
    L = L^* \left[ 1 - \frac{q_{1}}{1-x_\mathrm{i}(t)}\right]^{q_2},
\end{equation}
\begin{equation}
    T_\mathrm{f} = T_{\mathrm{f}}^* \left[1-\frac{q_3}{1-x_\mathrm{i}(t)} \right]^{q_4},
\end{equation}
where $q_{1},q_{2},q_{3},q_{4}$ are fitting constants that control the shape of the hyperbolic function and $f(x_i)$ reflects the dependence on the interface position. The fitting constants can be chosen to closely resemble the size-dependent relationship introduced into the Stefan condition in Section 2.2.4 ($q_1,q_3 = 10^{-3}, q_2,q_4 = 1$)


\section{Model verification and validation}
In the following analysis, various numerical simulations are developed with the objective of verifying the perturbation series solutions. Parametric studies are performed over the three key model parameters in the dimensionless system: the Stefan number ($\mathrm{Ste}$), the dimensionless thermal relaxation time ($\tilde \tau$), and the thermal diffusivity ($\alpha$), which together control the entire behavior of the dimensionless solution.

Convergence tests over the mesh size and the Courant number are presented in Figures 2 (a) and 2 (b) respectively, confirming numerical stability across varying spatial and temporal discretizations. Mesh independence and Courant number stability were validated for different values of the thermal relaxation time and Stefan number.

\begin{figure}[H]
    \centering
    \begin{subfigure}[t]{0.48\textwidth}
        \centering
        \includegraphics[width=\textwidth]{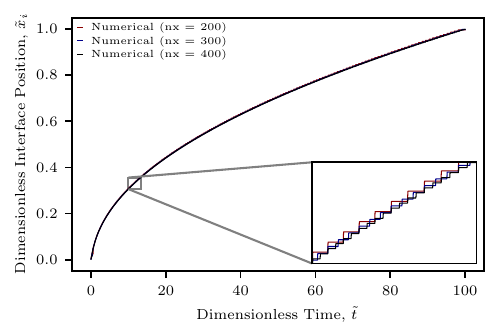}
        \caption{}
    \end{subfigure}
    \hfill
    \begin{subfigure}[t]{0.48\textwidth}
        \centering
        \includegraphics[width=\textwidth]{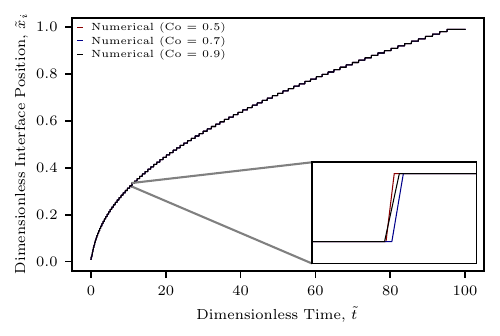}
        \caption{}
    \end{subfigure}
    \caption{Dimensionless interface position versus dimensionless time for (a) different mesh sizes ($\alpha = 1.0$, $\mathrm{Ste}=0.01$, $\tau=1.0$) and (b) different Courant numbers ($\alpha = 1.0$, $\mathrm{Ste}=0.01$, $\tau=1.0$).}
    \label{fig:mesh_courant_side_by_side}
\end{figure}

Numerical oscillations are most intense at high thermal relaxation times due to complications in resolving larger discrete interface jumps. Employing finer time and space discretizations can mitigate these numerical disturbances; however, mesh refinement alone proves insufficient when considering finite thermal-wave propagation speed. A numerical improvement over previous similar studies is achieved through the integration of the selective artificial viscosity outlined in Section 2.3 to suppress these remaining oscillations. Figure 3 demonstrates dimensionless temperature profiles taken under similar system conditions to those used in Glass et al. \cite{glass1990hyperbolic}, demonstrating a reduction in non-physical oscillations and improved resolution of the melt-front.
\begin{figure}[H]
    \centering
    \begin{subfigure}[t]{0.48\textwidth}
        \raggedright
        \textbf{(a)}\\
        \includegraphics[width=\textwidth]{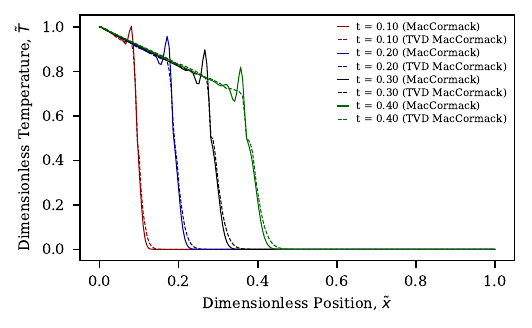}
    \end{subfigure}
    \hfill
    \begin{subfigure}[t]{0.48\textwidth}
        \raggedright
        \textbf{(b)}\\
        \includegraphics[width=\textwidth]{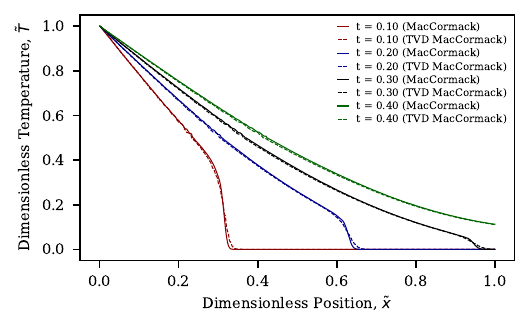}
    \end{subfigure}
    \caption{Dimensionless Temperature vs. Dimensionless Position for MacCormack--TVD and Standard MacCormack Methods for (a) high thermal relaxation time ($\alpha = 1.0$, $\mathrm{Ste}=10.0$, $\tau=1.0$) and (b) low thermal relaxation time ($\alpha = 1.0$, $\mathrm{Ste}=10.0$, $\tau=0.1$).}
    \label{fig:tvd_comp_tau01}
\end{figure}

Artificial viscosity suppresses spurious numerical oscillations near the phase interface, but introduces numerical diffusion that smears sharp solution features. A control parameter, $\lambda$, is therefore required to balance oscillation control against numerical resolution of the thermal and melting wave fronts. At lower Stefan numbers, the wave fronts become more discrete, and $\lambda$ must be reduced accordingly to prevent excessive smearing of these intense discontinuities.

\section{Results and Discussion}

In the following numerical simulations, we investigate a system subject to the following boundary and initial conditions:

\begin{align}
\tilde{T} = \tilde{T}_w, \quad \tilde{x} = 0, \quad \tilde{t} > 0 \\
\tilde{T} = \tilde{T}_0, \quad \tilde{x} \to X, \quad \tilde{t} > 0\\
\tilde{T} = \tilde{T}_0, \quad \tilde{x} \ge0, \quad \tilde{t} = 0
\end{align}
\begin{align}
\frac{\partial{Q}}{\partial{\tilde{x}}} = 0, \quad \tilde{x} = 0, \quad \tilde{t} > 0 \\
\frac{\partial{Q}}{\partial{\tilde{x}}} = 0, \quad \tilde{x} \to X, \quad \tilde{t} > 0\\
Q = 0, \quad \tilde{x} \ge0, \quad \tilde{t} = 0
\end{align}

The simulations presented in this section indicate strong agreement with both the composite and size-dependent latent heat models, evidenced by minimal error as the Stefan number approaches zero. The largest early-time contribution to error is the negative-time phenomenon in the size dependent latent heat model, discussed in Section 2.2.4, while the ``fast'' rescaling in the composite solution resolves the initial relaxation period with better model alignment (Figure 4). Early-time discrepancies are also partially attributable to numerical challenges in accurately resolving the discontinuous initial condition without generating non-physical artifacts. The amplitude of the oscillations can be reduced  by adding a non-physical smoothing boundary condition to avoid initially asymptotic behavior. Conversely, later-time deviations primarily stem from the finite truncation of the perturbation series. This error tends to scale proportionally to the lowest omitted order in the perturbation series, and variation from these theoretically predictable trends can be attributed to the potential divergent behavior of perturbation series. Therefore, while incorporating higher order terms could potentially reduce this truncation error, convergence is not assured, given that perturbation series often exhibit asymptotic rather than convergent power series \cite{hinch1991perturbation}.

The resulting interface profiles exhibit a characteristic parabolic-like shape, consistent with typical Stefan problem findings, with more significant early-time hyperbolic effects at larger thermal relaxation times. These thermal wave effects, prominent at early times, are a response to an augmented energy influx reflecting the influence of the thermal memory effects associated with the heat carriers. Once the thermal wave front has propagated through the entire spatial domain, the relaxation effects are no longer relevant, and the system transitions back to a classical diffusive-like framework.

\begin{figure}[H]
    \centering
    \includegraphics[width=0.75\linewidth]{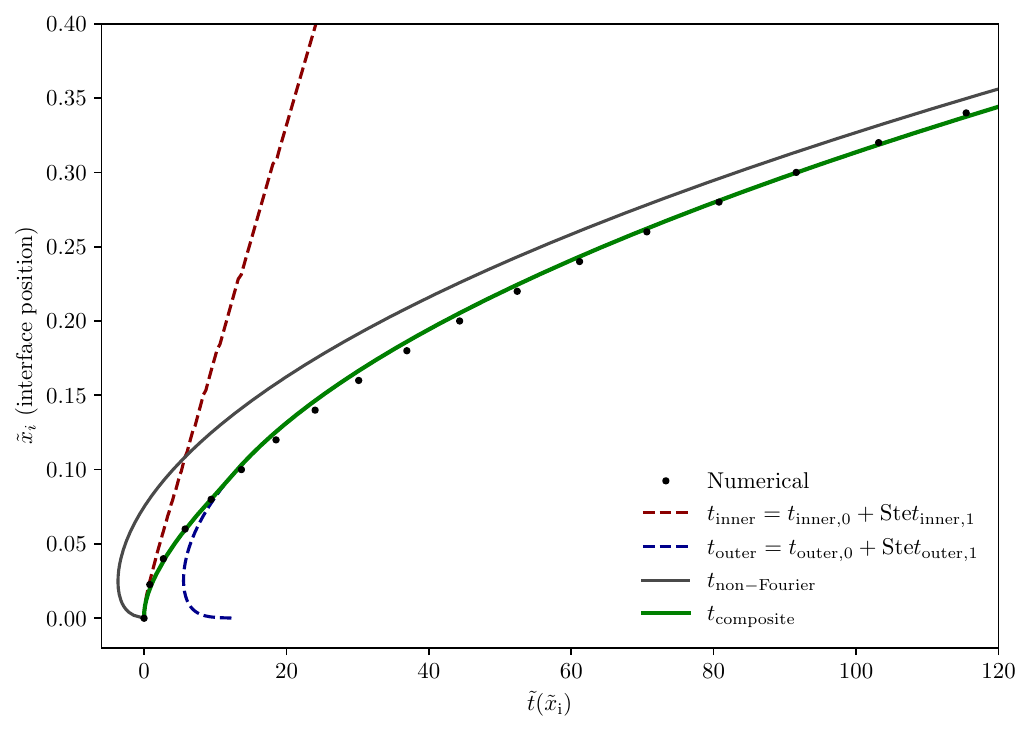}
    \caption{Numerical verification of composite and non-Fourier solution profiles for diffusive Stefan at small time-scales ($\alpha=1.0$, Ste = 0.001, $\tau = 1.0$)}
    \label{fig:comp_vs_numerical}
\end{figure}

\subsection{Effects of the Stefan Number}

\begin{figure}[H]
    \centering
    \begin{subfigure}[t]{0.48\textwidth}
        \raggedright
        \textbf{(a)}\\
        \includegraphics[width=\textwidth]{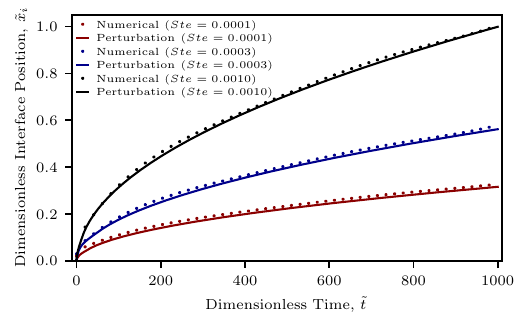}
        \label{fig:stefan_tau0_1}
    \end{subfigure}%
    \hfill
    \begin{subfigure}[t]{0.48\textwidth}
        \raggedright
        \textbf{(b)}\\
        \includegraphics[width=\textwidth]{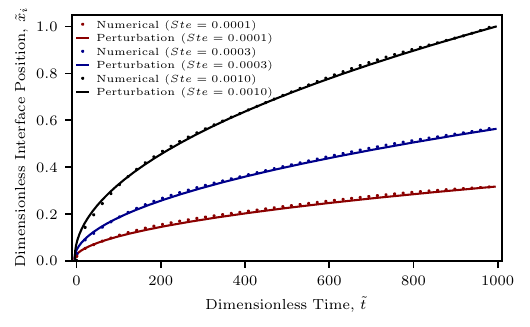}
        \label{fig:stefan_tau0_1_sizedep}
    \end{subfigure}
    \caption{Dimensionless Interface Position vs. Dimensionless Time for Different Stefan Numbers under the (a) Composite Solution with Diffusive Interface Condition ($\alpha = 1.0$, $\widetilde \tau = 1.0$) and (b) Size-Dependent, Non-Diffusive Interface Condition ($\alpha = 1.0$, $\widetilde \tau = 1.0$).}
    \label{fig:stefan_side_by_side}
\end{figure}

The Stefan number characterizes the ratio of sensible to latent heat throughout the phase change process. When the Stefan number is small (Ste $\ll$ 1), the latent heat component in the energy balance dominates the rate of interface motion:  for fixed $\tilde \tau$ and $\alpha$, the interface speed  decreases with the Stefan number (this is evident numerically when considering the dimensionalization of the characteristic time variable, which is inversely proportional to the Stefan number). Conversely, larger Stefan numbers indicate that sensible heat becomes increasingly limiting, which in the dimensionless system will correspond to more rapid interface progression. This trend is illustrated in Figures \ref{fig:stefan_side_by_side} (a) and~\ref{fig:stefan_side_by_side} (b).

The Stefan number also plays a central analytical role as the  perturbation parameter in the asymptotic series expansion. As $\mathrm{Ste} \to 0$, the leading-order temperature profile becomes quasi-steady, recovering the classical linear temperature distribution in the solid phase, admitting an exact closed-form solution at the leading order. Higher-order terms correct for departures from this idealized behavior; however, since the expansion is truncated at first order, the solutions cannot resolve dynamics at larger Stefan numbers, and the model error grows in proportion to the first omitted order of the expansion.

The inclusion of size dependent thermophysical values further modifies the interface profile. In finite systems, size-dependent thermophysical properties modify the effective latent heat at the phase-change interface, altering the early-time interface dynamics. This is illustrated in Figure~\ref{fig:stefan_side_by_side} (b), where the non-diffusive interface condition decreases the initial interface velocity which better captures the early-time hyperbolic behavior compared to the outer solution of the diffusive interface framework. Moreover, the relative error of the perturbation solution under the non-diffusive interface condition remains more uniform across the phase-change domain compared to the diffusive case, indicating a more physically faithful representation of the interface dynamics.

\subsection{Effects of the Thermal Relaxation time}

The second dimensionless parameter is the thermal relaxation time, $\widetilde \tau$, representing the intertemporal delay between an imposed temperature gradient and the establishment of the corresponding heat flux. In the classical framework, a thermal relaxation time of zero is assumed, which removes the hyperbolic terms in the MCV equation, recovering the classical heat equation. Analytically, this parabolic limit is approximately recovered in the small-$\mathrm{Ste}$ regime: the hyperbolic term in the dimensionless system (Eq. 5) carries a prefactor proportional to the Stefan number, making hyperbolic contributions to the interface profile amenable to a perturbation series expansion.

Physically, $\widetilde \tau =0$ implies instantaneous adjustments of the heat flux in response to an imposed temperature gradient. However, in the hyperbolic formulation, the heat carriers require a characteristic finite time to respond to thermal perturbations, so that energy initially propagates as damped thermal waves rather than by purely diffusive spreading. Larger values of $\widetilde \tau$ prolong these wave-like effects, and thus the delayed onset of the interface motion is a consequence of the early-time relaxation effects. 

\begin{figure}[H]
    \centering
    \begin{subfigure}[t]{0.48\textwidth}
        \raggedright
        \textbf{(a)}\\
        \includegraphics[width=\textwidth]{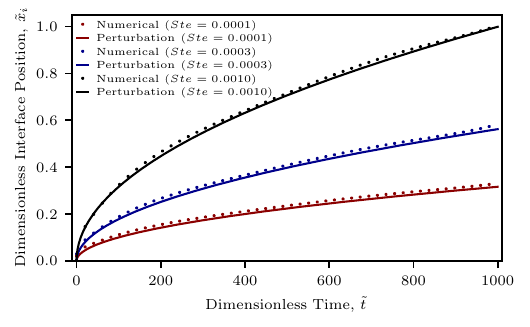}
        \label{fig:alpha_tau01}
    \end{subfigure}%
    \hfill
    \begin{subfigure}[t]{0.48\textwidth}
        \raggedright
        \textbf{(b)}\\
        \includegraphics[width=\textwidth]{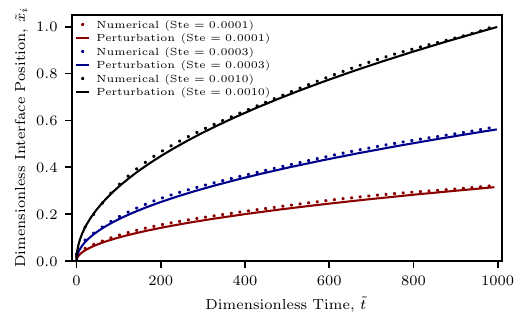}
        \label{fig:alpha_tau1}
    \end{subfigure}
    \caption{Dimensionless Interface Position vs. Dimensionless Time for Different Stefan Numbers approaching the parabolic limit under the (a) Composite Solution with Diffusive Interface Condition ($\alpha = 1.0$, $\widetilde \tau = 0.01$) and (b) Size-Dependent, Non-Diffusive Interface Condition ($\alpha = 1.0$, $\widetilde \tau = 0.01$)}
    \label{fig:tau_parametric_study}
\end{figure}

This is shown in Figure \ref{fig:tau_parametric_study} which shows a similar study to Figure 5, except approaching the parabolic limit ($\widetilde \tau \to0)$. In the small-$\widetilde \tau$ simulation, the smaller delay in heat flux response increases the amount of thermal energy that initially reaches the phase change front compared to more hyperbolic simulations. Consequently, the initial energy available to satisfy the Stefan condition is increased, speeding the early-time interface propagation relative to the less diffusive model. Solutions at smaller $\widetilde \tau$ also exhibit reduced model error, concentrated primarily at early times, as seen in the improved agreement between the perturbation solution and the numerical result as the limit of $\widetilde \tau \to 0$ is approached in the non-diffusive interface formulation (Figures 5 (b) and 6 (b)). The Fourier interface model, by contrast, shows no significant sensitivity to this limit, as the composite solution already captures these dynamics through the matched inner layer.

\subsection{Effects of the Thermal Diffusivity}
\begin{figure}[H]
    \centering
    \begin{subfigure}[t]{0.48\textwidth}
        \raggedright
        \textbf{(a)}\\
        \includegraphics[width=\textwidth]{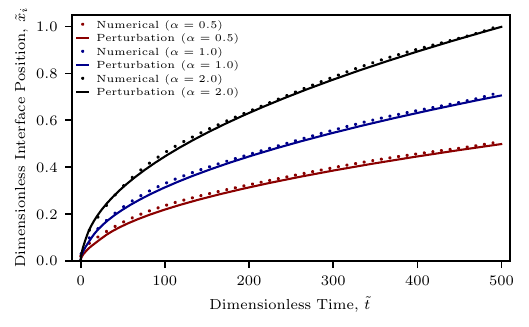}
    \end{subfigure}%
    \hfill
    \begin{subfigure}[t]{0.48\textwidth}
        \raggedright
        \textbf{(b)}\\
        \includegraphics[width=\textwidth]{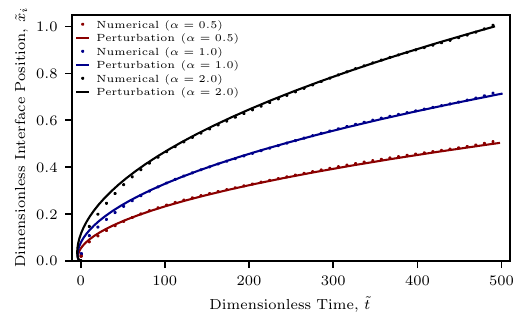}
    \end{subfigure}
    \caption{Dimensionless Interface Position vs. Dimensionless Time for for Different Thermal Diffusivities ($\alpha$) under the (a) Composite Solution with Diffusive Interface Condition (Ste = 0.001, $\widetilde \tau$ = 1.0) (b) Size-Dependent, Non-Diffusive Interface Condition (Ste = 0.001, $\widetilde \tau$ = 1.0).}\label{fig:alpha_param}
\end{figure}

Contrary to the Stefan number and the dimensionless thermal relaxation time, the thermal diffusivity, $\alpha$, does not appear in the governing system. However, it enters through the non-dimensionalization of both the time variable and the thermal relaxation time, and so acts as a scaling parameter on the temporal and hyperbolic structure of the problem. Physically, the thermal diffusivity represents the rate at which temperature disturbances spread through a material. Materials with higher diffusivity transport thermal energy more efficiently, allowing heat to reach the solid–liquid interface more rapidly. Consequently, the energy required for phase change is supplied more quickly in systems with larger $\alpha$, resulting in faster advancement of the phase change front. In the hyperbolic formulation, $\alpha$ also sets the characteristic thermal wave speed,
\begin{equation}
\textit{c} = \sqrt{\frac{\alpha}{\tau}},
\end{equation}
so that $\textit{c}$ scales as $\sqrt{\alpha}$, and the total phase-change time scale decreases as 1/$\alpha$. This behavior is consistent with the similarity solution to the parabolic Stefan problem, where the interface position scales with $\sqrt{\alpha t}$ \cite{kakacc2018heat}. As a result, the relative length of the domain in which relaxation effects are relevant increases with $\alpha$, an effect that can be seen in Figure \ref{fig:alpha_param}. 

The negative early-time divergence visible in Figure 7 (b) arises from the structure of the perturbation expansion when applied to the non-diffusive Stefan condition. In the diffusive interface formulation, hyperbolic corrections enter only through the temperature profile; in the non-diffusive formulation, they appear additionally in the Stefan condition itself. These compounded contributions produce a stronger correction to the early-time dynamics, amplifying the negative-time divergence discussed in Section 2.2.4. Since increasing $\alpha$ increases the relative domain in which hyperbolic effects are relevant, the amplification of these negative-time artifacts grows with $\alpha$, an effect that is attributed to the low-order approximate nature of the perturbation series solution.

\section{Conclusion}
This work presents an analytical and numerical framework for the one-dimensional hyperbolic Stefan problem governed by the MCV equation in both diffusive and non-diffusive Stefan condition frameworks. The non-diffusive interface condition behavior was approximated with the introduction of size-dependent thermophysical values (melting temperature and latent heat of phase change). The systems were analyzed in the small limit of the Stefan number, enabling a perturbation series approximation used to capture the interface behavior.

The perturbation series solutions were initially constructed around the temperature profile and the $\tilde t(\widetilde{x}_{\mathrm{i}})$ reformulation in the diffusive Stefan condition formulation, yielding closed-form solutions for both temperature and interface location in time. The first-order solution for the interface position contains a logarithmic divergence as $\widetilde{x}_{\mathrm{i}} \to 0^+$, which fails to capture the initial condition. To fix this, two solution methods were proposed. Firstly, an inner solution derived on a rescaled time variable ($\phi = \widetilde{t}/\sqrt{\mathrm{Ste}}$) was introduced to capture the initial condition and early-time dynamics. This completed the previously derived "outer" solution with a smoothing function to a uniformly valid approximation across the entire interface domain.

The second proposed fix was to introduce non-Fourier heat transfer effects at the phase change interface in the way of size-dependent latent heat and melting temperature. The first order interface solution in this formulation also possesses a logarithmic divergence, but in the limit of $L^* \to\infty$ ($\equiv $Ste$\to 0$), a substitution is applied to obtain a closed-form, well-defined solution over the full interface domain. The resulting non-diffusive interface formulation solution initially predicts negative time values; an approximate artifact of the initial relaxation period present in hyperbolic heat transfer, and made more prominent in this formulation as the Stefan condition, which in this formulation is considered under a hyperbolic framework, is the governing equation used to obtain the interface profile solution. This initial artifact leads to better alignment with the numerical verification model, despite initially non-physical behavior.

The numerical verification of the models was done against a total variation diminishing MacCormack predictor-corrector scheme with selective artificial viscosity based on a TVD flux-limiter framework. The scheme obtains second-order accuracy in both space and time, and the selective artificial viscosity dampens the numerical Gibbs-type oscillations often present in hyperbolic free boundary problems at the thermal and phase change wave front.

Parametric studies were completed over three pertinent parameters in the dimensionless system: the Stefan number (Ste), the dimensionless thermal relaxation time ($\widetilde \tau$), and the thermal diffusivity ($\alpha$). The error of the diffusive interface framework scales differently in the inner and outer solutions; the inner solution error scales $\propto$ Ste, whereas the outer solution and the non-diffusive interface framework solution have errors that scale $\propto$ Ste$^2$. These error relationships are the truncation errors present in perturbation theory. Larger values of $\widetilde \tau$ amplify early-time hyperbolic effects, increasing the perturbation of the temperature and interface solution profiles from the idealized Ste = 0 base case. This results in a less accurate capture of the higher order corrective terms in the series solutions and higher model error. Larger $\alpha$ increases thermal wave speed and extends the fraction of the total temporal domain in which hyperbolic behavior dominates, which in turn increases the model error.

Future work could extend the framework to different geometries, particularly spherical and cylindrical, relevant for nanoscale melting processes like droplet freezing or cylindrical phase change materials. A two-phase solution model could be considered, allowing analysis of systems in which the second phase is not isothermal and contributes a non-trivial heat flux to the interface energy balance. This would also enable the study of systems where the two phases have different thermophysical properties, such as differing thermal relaxation times or thermal diffusivities \cite{Liu2009Hyperbolic}.

\section*{Conflict of Interest}
The authors declared that there is no conflict of interest.

\section*{Data Availability}
Data will be made available on request.

\section*{CRediT authorship contribution statement}
Conceptualization: SH and MX.
Methodology: MVH, MX, SH.
Software: MVH and SH.
Validation: MVH, MX, SH.
Formal analysis: MVH and MX.
Investigation: MVH, MX, SH.
Resources: SH.
Data curation: MVH.
Writing - original draft preparation: MVH and MX.
Writing - review and editing: SH.
Visualization: MVH.
Supervision: SH and MX.
Project administration: SH and MX.
Funding acquisition: SH.

\clearpage
\section*{Nomenclature}

\begin{tabbing}
\hspace{2.8cm} \= \kill
\textbf{Roman symbols} \> \\[2pt]
$A_n$, $B_n$       \> Cosine and sine modal coefficients in the inner temperature expansion \\
$b_n$              \> Modal ODE coefficient in the inner temperature expansion \\
$B$                \> Jacobian matrix in the MacCormack scheme \\
$c$                \> Thermal wave speed ($\mathrm{m\,s^{-1}}$) \\
$C_n$              \> First-order modal coefficients in the inner temperature expansion \\
$c_p$              \> Specific heat capacity ($\mathrm{J\,kg^{-1}\,K^{-1}}$) \\
$C(\cdot)$          \> Local Courant number function \\
$D_{i\pm1/2}$      \> Numerical flux difference terms in the TVD artificial viscosity \\
$F$                \> Conservative flux vector in the MacCormack scheme \\
$G_i^{\pm}$        \> Numerical viscosity coefficients in the TVD scheme \\
$H$                \> Enthalpy ($\mathrm{J}$) \\
$k$                \> Thermal conductivity ($\mathrm{W\,m^{-1}\,K^{-1}}$) \\
$l$, $w$           \> Material slab thickness ($\mathrm{m}$), and width ($\mathrm{m}$) \\
$L$, $L^*$         \> Size-dependent and bulk latent heat of fusion ($\mathrm{J\,kg^{-1}}$) \\
$q$                \> Heat flux in the enthalpy formulation ($\mathrm{W\,m^{-2}}$) \\
$q_1, q_2, q_3, q_4$ \> Fitting constants in the size-dependent thermophysical functions \\
$Q$                \> Conservative heat flux variable in the numerical formulation ($\mathrm{W\,m^{-2}}$) \\
$r_j^{\pm}$        \> Forward/backward cell gradient ratios in the flux limiter \\
$S(U)$             \> Source term vector in the numerical formulation \\
$\mathrm{Ste}$     \> Stefan number \\
$t$, $\tilde{t}$   \> Time ($\mathrm{s}$), and its dimensionless equivalent \\
$T$                \> Temperature ($\mathrm{K}$) \\
$T_{\mathrm{c}}$   \> Cold-wall temperature ($\mathrm{K}$) \\
$T_{\mathrm{f}}$, $T_{\mathrm{f}}^*$ \> Phase change and bulk phase change temperature ($\mathrm{K}$) \\
$T_m^{\pm}$        \> Upper/lower bounds of the enthalpy smoothing interval ($\mathrm{K}$) \\
$\widetilde{T}$    \> Dimensionless temperature \\
$U$                \> Conservative solution vector in the numerical scheme \\
$u(\cdot)$ \> Blending function in the composite interface approximation \\
$v(\cdot, \cdot)$ \> Homogeneous correction in the inner temperature expansion \\
$x$, $X$           \> Spatial coordinate ($\mathrm{m}$), and domain length ($\mathrm{m}$) \\
$x_\mathrm{i}$, $\tilde{x}_\mathrm{i}$ \> Interface position ($\mathrm{m}$), and its dimensionless equivalent \\
$\tilde{x}$        \> Dimensionless spatial coordinate \\
$x_t$, $\gamma$    \> Total slab length ($\mathrm{m}$), and molten sub-layer length ($\mathrm{m}$) \\
$\delta V$         \> Differential volume of the molten surface sub-layer ($\mathrm{m^3}$) \\[6pt]
\textbf{Greek symbols} \> \\[2pt]
$\alpha$           \> Thermal diffusivity ($\mathrm{m^2\,s^{-1}}$) \\
$\delta(\cdot,\cdot)$    \> Regularized Dirac delta function in the enthalpy formulation \\
$\epsilon$         \> Smoothing parameter in the enthalpy method \\
$\lambda$          \> Artificial-viscosity control parameter \\
$\mu(\cdot)$           \> Artificial viscosity coefficient \\
$\nu_i$            \> Local Courant number \\
$\theta(\cdot)$           \> Flux limiter function \\
$\phi$             \> Inner fast-time variable \\
$\psi(\cdot,\cdot)$             \> Sensible heat contribution in the enthalpy formulation \\
$\rho$             \> Mass density ($\mathrm{kg\,m^{-3}}$) \\
$\tau$, $\tilde{\tau}$ \> Thermal relaxation time ($\mathrm{s}$), and its dimensionless equivalent \\
$\omega_n$         \> Modal frequency
\\[6pt]
\textbf{Superscripts and subscripts} \> \\[2pt]
$0$                \> Leading-order solution \\
$1$                \> First-order solution \\
$\mathrm{c}$       \> Cold-wall quantity \\
comp               \> Composite solution \\
f                  \> Phase change or fusion state \\
i                  \> Interface quantity \\
in, out            \> Inner and outer solution \\
$m$                \> Melting quantity \\
$n$                \> Mode index or time level (by context) \\
\end{tabbing}
\clearpage
 \bibliographystyle{elsarticle-num} 
 \bibliography{REF}

\end{document}